\documentclass{aa}  

\usepackage{subcaption}

\usepackage{graphicx}

\usepackage{amsthm,amssymb,amsmath}
\usepackage{siunitx}
\DeclareSIUnit\angstrom{\text{Å}}
\usepackage{txfonts}
\usepackage{color}
\usepackage{booktabs}
\usepackage{array}
\usepackage{multirow}
\usepackage{natbib}
\usepackage{geometry}
\usepackage{soul}
\usepackage{orcidlink}

\bibpunct{(}{)}{;}{a}{}{,} 
\definecolor{cobalt}{rgb}{0.06, 0.2, 0.65}
\hypersetup{
  colorlinks,
  citecolor=cobalt,
  linkcolor=[rgb]{0.8, 0.2, 1.0},
  urlcolor=cobalt,
}

\begin{document}

   \title{The environments of radio galaxies and quasars in LoTSS data release 2}

 \author{Tong Pan\orcidlink{0000-0002-4825-6131} \inst{1}
         \and
         Yuming Fu\orcidlink{0000-0002-0759-0504}\inst{1,2}
         \and
          H.~J.~A.~Rottgering\inst{1}
        \and
        R.~J.~van Weeren \inst{1}
        \and
        A.~B.~Drake \inst{3}
        \and
        B.~H.~Yue\inst{1,4}
        \and
        J.~W.~Petley\inst{1}
        }

  \institute{Leiden Observatory, Leiden University, P.O. Box 9513, NL-2300 RA Leiden, The Netherlands\\
              \email{tpan@strw.leidenuniv.nl}
             \and Kapteyn Astronomical Institute, University of Groningen, P.O. Box 800, NL-9700 AV Groningen, The Netherlands 
             \and Centre for Astrophysics Research, University of Hertfordshire, Hatfield AL10 9AB, UK 
             \and Institute for Astronomy, University of Edinburgh, Edinburgh EH9 3HJ, UK 
            }



   \abstract
   {} 
   {The orientation-based unification scheme of radio-loud active galactic nuclei (AGNs) asserts that radio galaxies and quasars are essentially the same type of object, but viewed from different angles. To test this unification model, we compared the environments of radio galaxies and quasars, which would reveal similar properties when an accurate model is utilized.}
   {Using the second data release of the LOFAR Two-metre Sky Survey (LoTSS DR2), we constructed a sample of 26,577 radio galaxies and 2028 quasars at $0.08 < z < 0.4$. For radio galaxies with optical spectra, we further classified them as 3631 low-excitation radio galaxies (LERGs) and 1143 high-excitation radio galaxies (HERGs). We crossmatched these samples with two galaxy cluster catalogs from the Sloan Digital Sky Survey (SDSS).}
   {We find that $17.1 \pm 0.2 \%$ of the radio galaxies and $4.1 \pm 0.4 \%$ of the quasars are associated with galaxy clusters. Luminous quasars are very rare in clusters, while $18.7 \pm 0.7 \%$ LERGs and $15.2 \pm 1.1 \%$ HERGs reside in clusters. We also note that in radio galaxies, both HERGs and LERGs tend to reside in the centers of clusters, while quasars do not show a strong preference for their positions in clusters.}
   {This study shows that local quasars and radio galaxies exist in different environments, challenging the orientation-based unification model. This means that factors other than orientation may play an important role in distinguishing radio galaxies from quasars. The future WEAVE-LOFAR survey will offer high-quality spectroscopic data for a large number of radio sources and allow for a more comprehensive exploration of the environments of radio galaxies and quasars.}


   \keywords{galaxies: active --
                quasars: general --
                galaxies: clusters: general --
                radio continuum: galaxies
               }

   \maketitle
%

\section{Introduction}
     
    According to the orientation-based unified model of radio-loud active galactic nuclei (AGNs), quasars and radio galaxies are essentially the same radio sources, differing only in their appearance based on the observer's line of sight \citep{1989barthel,Antonucci&miller1985,Cimatti1993,Urry&Paolo1995,Gopal-Krishna1995}. This unified model asserts that both radio galaxies and radio-loud quasars have a supermassive black hole, an accretion disk, a broad line region (BLR), a dusty torus, a narrow line region (NLR), and radio jets. When the observer's line of sight falls within the opening angle of the torus, the nucleus continuum and BLR are unobscured, and the source is seen as a quasar with broad emission lines. When the observer's line of sight passes through the dusty torus, the nucleus continuum and BLR are obscured, and the source is seen as a radio galaxy with narrow emission lines.

    One prediction of the orientation-based unification scheme is that radio galaxies have larger projected linear sizes in radio than quasars do. Several studies have obtained results that are consistent with this prediction, demonstrating that the linear sizes of radio galaxies are (on average) larger than those of quasars \citep[e.g.,][]{1989barthel,singal2014,2017Morabito}.

    The orientation-based unified model also predicts that radio galaxies and quasars should have similar environments. However, differences have been observed between the environments of radio galaxies and those of quasars. For example, radio galaxies are typically found in massive, elliptical galaxies \citep[see e.g.,][]{kauffmann2003,2009hickox}, while quasars are found in a wider range of galaxy types, including spiral, elliptical, and merging galaxies \citep[see e.g.,][]{2003dunlop,Bahcall1997ApJ,1999mclure,2016kellermann}. In addition, radio galaxies are more likely to live in higher density environments than quasars from the cluster scale to the super-cluster scale, as revealed by clustering and luminosity-density field analyses \citep{Donoso2010MN,Lietzen2011aa}. One possible explanation is that radio galaxies and quasars might have been formed in dark matter halos of different masses.

    \cite{2001harvanek} compared the environments of 66 radio galaxies and 14 quasars and found that radio galaxies are much more likely to be located in galaxy clusters, as compared to quasars. However, their results were questioned by \cite{2004Hardcastle} due to the inclusion of low-excitation radio galaxies (LERGs) in the radio galaxy sample. The subtype of radio galaxies known as LERGs exhibit weak or absent emission lines and are thought to be powered by the ``hot-mode'' accretion, where hot gas is accreted via radiatively inefficient flows \citep[e.g.,][]{1994narayan&li,2008ho,2012best&heckman}. As a result, LERGs are typically found in high-density environments (e.g., galaxy clusters), where the hot gas suppresses their accretion efficiency. In contrast, high-excitation radio galaxies (HERGs), with strong optical emission lines, are powered by  ``cold-mode'' accretion, where the cold gas is accreted efficiently through a radiatively efficient thin disk \citep{2007hardcastle, 2012best&heckman,2022thomas}. Therefore, when studying the environments of radio galaxies, it is essential to account for these different subclasses, as they are typically found in distinct environments.

    In previous studies, samples of radio galaxies and quasars typically came from small low-frequency catalogs, such as the Third Cambridge Catalogue \citep[3C;][]{1959edge,1962bennett} and the Seventh Cambridge Redshift Survey \citep[7C;][]{1999lacy,2004grims}. These studies have also utilized high-frequency surveys like the NRAO VLA Sky Survey \citep[NVSS;][]{1998AJ....115.1693C}, which are generally less sensitive to diffuse emission.

    To overcome these limitations, we chose to  utilize data from the Low-Frequency Array \citep[LOFAR;][]{2013A&A...556A...2V}, which operates at frequencies between 10 and 240 MHz. We have systematically investigated the environments of radio galaxies and quasars by selecting highly complete samples of both classes from the LOFAR Two-Metre Sky Survey \citep[LoTSS;][]{shimwell2017}. LoTSS has a significant advantage in selecting quasars and radio galaxies in an orientation-independent manner thanks to the fact that the low-frequency observations preferentially detect extended emission \citep[see, e.g.,][]{2020A&A...642A.107C}, whose morphologies are not strongly related to the orientations of the galaxies.

    We use the optically identified catalog of LoTSS DR2 \citep{hardcastle2023} to select our sources, which allows for reliable statistical studies to be undertaken. We primarily followed the selection procedures outlined by \cite{2019hardcastle} to construct samples of radio galaxies and quasars and further classify the radio galaxies with emission-line data into LERGs and HERGs. By crossmatching the radio galaxy and quasar samples with catalogs of galaxy clusters, we obtained the fractions of radio galaxies and quasars
    that are associated with clusters. These fractions, as well as the distances between radio galaxies and quasars and the cluster centers, enable us to carry out quantitative comparisons of the environments of radio galaxies and quasars.

    The structure of this paper is as follows. Section 2 describes the selection procedures of radio galaxies and quasars and the crossmatch method for identifying their associations with galaxy clusters. Section 3 shows the results of significantly different environments of radio galaxies and quasars, as revealed by the fractions of radio galaxies and quasars associated with clusters. Section 4 discusses the possible reason for the environmental differences between radio galaxies and quasars. Throughout the paper, we assume a flat $\Lambda$CDM cosmology with $H_0 = 70~\mathrm{km~s^{-1}~Mpc^{-1}}$ and $\Omega_{M}=0.3$.

\section{Samples and method}
\subsection{The radio catalog}

    The LOFAR Two-Metre Sky Survey \citep[LoTSS;][]{shimwell2017} is a high-angular-resolution survey conducted by LOFAR, aimed at mapping the entire northern hemisphere within the frequency range of 120 to 168 MHz.
    The second data release of LoTSS \citep[LoTSS-DR2;][]{2022shimwell} covers 27\% of the northern sky and is divided into two regions centered at 12h45m00s +44\(^\circ 30'00''\) and 01h00m00s +28\(^\circ 00'00''\), spanning 4178 and 1457 square degrees, respectively. The LoTSS-DR2 images have a median rms noise of 83 $\mu$Jy/beam at a resolution of 6${''}$. The catalog comprises 4,396,228 sources. For 85\% (4,167,359) of the LoTSS-DR2 sources, \citet[][hereafter H23]{hardcastle2023} have provided their optical counterparts from the DESI Legacy Imaging Survey \citep{2019dey} and mid-IR counterparts from unWISE \citep{2019Schlafly}. The radio galaxies and quasars used in this work are selected from the H23 sample (Section \ref{sec2_3}).

\subsection{Galaxy cluster catalogs}
\label{sec_cluster}

    We aim to use galaxy cluster catalogs to probe whether the environments of radio galaxies and quasars are different. We follow \cite{2019croston} who used two cluster catalogs both derived from Sloan Digital Sky Survey Data Release 8 \citep[SDSS DR8;][]{2011Aihara}. These two catalogs are: the RedMaPPer catalog \citet[][referred to as R14]{2014rykoff}, and the galaxy cluster catalog by \citet[][referred to as W12]{2012wen}. R14 covers the redshift range of $0<z<0.944$ and has over 25,000 clusters selected through a red-sequence finding method, among which approximately 4000 clusters are within the LoTSS DR2 footprint. W12 covers a redshift range of $0.05<z<0.78$ and has more than 130,000 clusters. These clusters are selected with an iterative approach that combines photometric redshift selection with a friends-of-friends (FoF) method. Among all W12 clusters, around 9600 are within the LoTSS DR2 area. R14 and W12 have comparatively well-calibrated richness estimators $\lambda$ and $R{_{L^*}}$, with W12 extending to a lower richness than R14. Following \cite{2019croston}, we impose a redshift range of $0.08 < z < 0.4$, resulting in approximately 2600 clusters in R14 and 8000 clusters in W12 within the LoTSS DR2 area. The reason for setting this redshift range is that R14 is $>85\%$ complete above $\lambda$ = 30 and $>95\%$ above $\lambda$ = 40 at $0.08 < z < 0.4$, while W12 is $>95\%$ complete up to $z=0.42$. 
    For a detailed discussion of these completeness properties, we refer the reader to \cite{2019croston}. In this redshift range, the $M_{200}$ of W12 clusters are $> 3.9 \times 10^{14} \, M_{\odot}$ and the $M_{200}$ of R14 clusters are $> 1.07 \times 10^{14} \, M_{\odot}$. Both catalogs are $> 95\%$ complete for $M_{200} > 10^{14} \, M_{\odot}$.
    In Section \ref{sec:result1}, we  demonstrate that the W12 catalog exhibits higher completeness compared to R14. Therefore, our primary findings have been derived from W12.


\subsection{Samples of radio galaxies and quasars} \label{sec2_3}

    To build samples of radio galaxies and quasars from H23, we apply a flux threshold of $>$ 1.1 mJy, at which the LoTSS DR2 is more than $95\%$ complete \citep{2022shimwell}, resulting in 1,776,977 radio sources in the H23 catalog. We then impose a redshift range of $0.08 < z < 0.4$ based on the completeness of clusters described in Section \ref{sec_cluster} and obtain 281,552 radio sources. Within this redshift range, 1010 sources in H23 have already been identified as quasars by crossmatching the LoTSS DR2 catalog with the 16th data release of the SDSS quasar catalog \citep[DR16Q;][]{2020sdssquasar}. 
    
\begin{figure}[htb]
    \centering
    \includegraphics[width=0.38\textwidth]{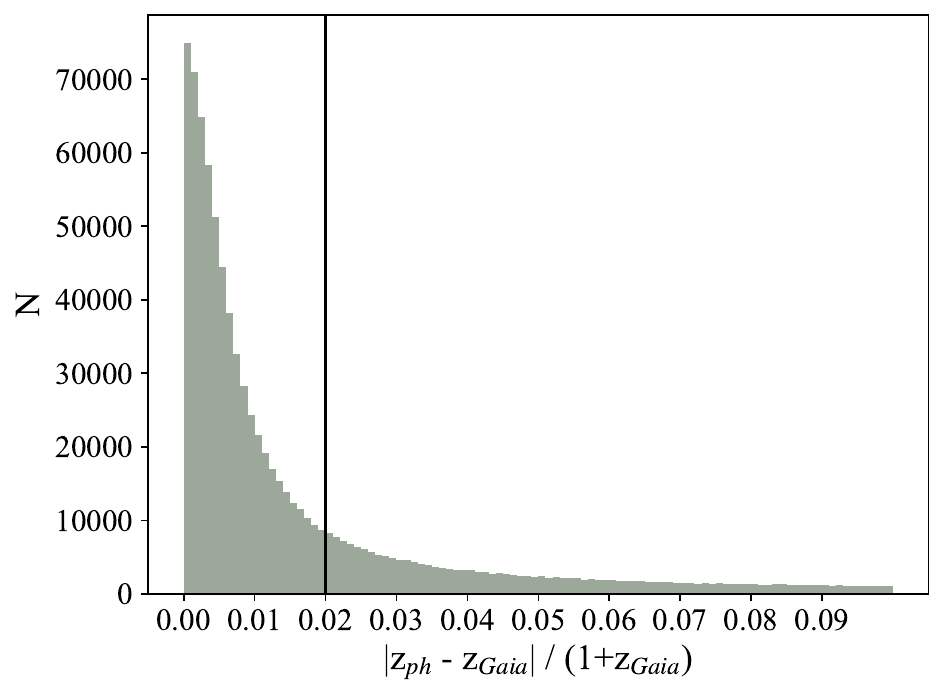}
    \caption{Histogram of $|z_{ph}$ - $z_{Gaia}| / (1+z_{Gaia})$.  Black vertical line marks $|z_{ph}$ - $z_{Gaia}| /(1+z_{Gaia})$ = 0.02. }
    \label{fig_deltaz}
\end{figure}

    To increase the sample size of the LoTSS quasars, we further select quasars from CatNorth \citep{2024Fu}, an improved Gaia DR3 \citep{Gaiadr3} quasar candidate catalog \citep{gaiadr3quasar} with more than 1.5 million sources in the 3$\pi$ sky built with data from Gaia, Pan-STARRS1 \citep{2016panstarrs}, and CatWISE2020 \citep{catwise2020}. CatNorth provides photometric redshifts ($z_{ph}$), which have a much lower fraction of outliers than the original Gaia redshifts ($z_{Gaia}$). Nevertheless, when the two redshifts estimates are close (e.g., $\Delta z / (1+z_{Gaia}) < 0.02$, where $\Delta z = |z_{ph} - z_{Gaia}|$), $z_{Gaia}$ has higher precision than $z_{{ph}}$ because the former is measured from the low-resolution spectroscopy of Gaia. Figure \ref{fig_deltaz} shows the histogram of $\Delta z /(1+z_{Gaia})$. We applied $\Delta z /(1+z_{Gaia}) < 0.02$ and $0.08 < z_{Gaia} < 0.4$ (based on the completeness of cluster catalogs) and obtain 22,719 quasars in CatNorth with precise Gaia redshifts. We then crossmatched them with the 1,776,977 radio sources above the flux threshold of 1.1 mJy in the H23 catalog using a 1.5-arcsec matching radius and obtain 1018 quasars that are not in DR16Q. In total, we obtain 2028 quasars, with 1010 having SDSS redshifts and others having Gaia redshifts.

    To create a sample of radio galaxies, we first exclude sources identified as quasars from above. We then exclude sources with Ks-band absolute magnitude \( \mathrm{M_{ks}} < -17 \) or \(\mathrm{M_{ks}} > -33, \) which would indicate unreliable redshift values \citep{2019hardcastle}. This leaves us with 262,857 radio sources. This population comprises both radio galaxies and star-forming galaxies (SFGs). We then applied various diagnostic methods suggested by \citet{2019sabeter} to distinguish SFGs from radio AGNs:

    \begin{enumerate}
        \item The BPT diagram \citep{baldwin1981,kauffmann2003,kewley2006}. Sources are classified as SFGs if they satisfy the condition 
        \(\log\left(\mathrm{[O~\textsc{iii}]}/{\mathrm{H{\beta}}}\right) > 1.3 + 0.61 (\log\left({\mathrm{[N~\textsc{ii}]}}/{\mathrm{H{\alpha}}}\right) -0.05). \)
        Sources that do not meet this condition are classified as radio AGNs.
        
        \item The \(D_{4000}\) versus \( L_{\mathrm{144MHz}}/{M_{\ast}} \) plane \citep{Best2005}. $D_{4000}$ is the strength of $\SI{4000}{\angstrom}$ break in the spectrum of a galaxy and $L_{\mathrm{144MHz}}/M_{\ast}$ is the ratio of the radio luminosity to stellar mass.
        Sources are classified as SFGs if they are below the second line reported in Figure 1 of \cite{2019sabeter}. Sources that do not meet this condition are classified as radio AGNs. 
        
        \item The $L_{\mathrm{H{\alpha}}}$ versus $L_{\mathrm{144MHz}}$ plane. Sources are classified as SFGs if they satisfy the condition
        \( \log({L_{\mathrm{H{\alpha}}}}/{L_{\ast}}) > \log(L_{\mathrm{144MHz}}) - 16.9 \). Sources that do not meet this condition are classified as radio AGNs.
        
        \item The WISE W2–W3 colour cut \citep{wright2010,mateos2012,2014Gurkan,herpich2016}. Sources are classified as SFGs if they satisfy the condition $\rm W2 - W3 > 0.8$ (AB) or, equivalently, $\rm W2 - W3 > 2.635$ in Vega magnitude. Sources that do not meet this condition are classified as radio AGNs.
        
    \end{enumerate}
    
    Diagnostic 1-3 require measurements of the optical spectra of galaxies. To get this information, we crossmatched the 262,857 radio sources with the MPA-JHU catalog of galaxy properties\footnote{\url{https://www.sdss4.org/dr17/spectro/galaxy_mpajhu/}} \citep{2004brinchmaan}, which is derived from the optical spectra of the SDSS DR8. There are 45,198 radio sources with optical counterparts in this catalog. Diagnostic 4 requires WISE data from the H23 catalog. We imposed magnitude error limits of less than 0.2 for W1 and W2, and less than 0.7 for W3 (corresponding to signal-to-noise ratios, S/N values, of 5 and 3, respectively). This selection yields 37,363 radio sources with emission-line data and unWISE colours. Following the combination method of the above four diagnostics in \cite{2019sabeter}, we identify 27,980 SFGs among the 37,363 radio sources, leaving 9383 sources classified as radio galaxies.

\begin{figure}[htb]
    \centering
    \includegraphics[width=0.38\textwidth]{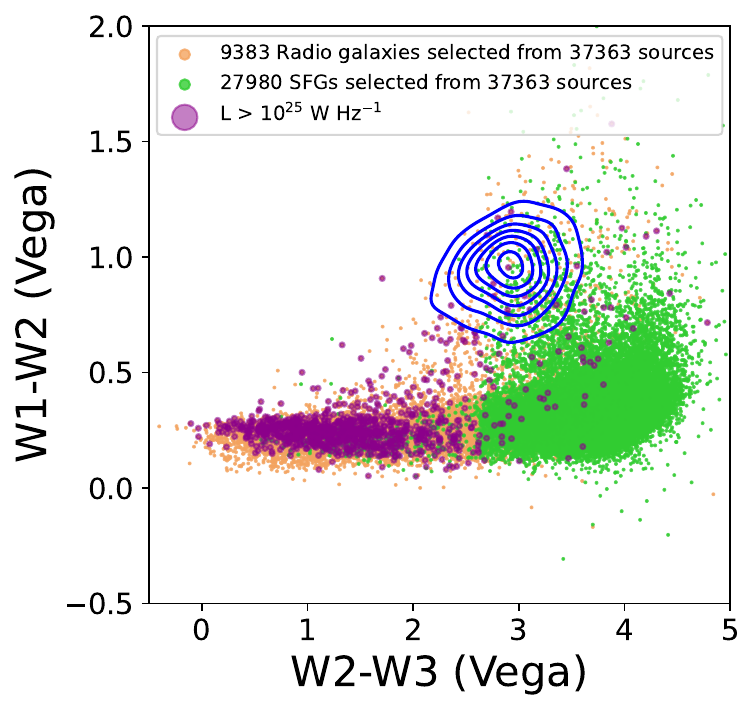}
    \caption{WISE colour–colour diagram of the 37,363 spectral sources from MPA-JHU catalog. Green dots, orange dots, and purple dots represent SFGs, radio galaxies, and luminous radio sources ($L>10^{25}~\mathrm{W~Hz^{-1}}$), respectively. The blue contour lines represent the density distribution of quasars. }
    \label{fig_mpa}
\end{figure}

    For the 217,659 radio sources that are not in the MPA-JHU catalog (and therefore lack emission-line data), we  determined their classes using WISE colours and radio luminosity. We retained 185,476 radio sources with available and accurate WISE colours (same magnitude error limits of less than 0.2 for W1 and W2, and less than 0.7 for W3). Since we have already classified the spectral samples as radio galaxies and SFGs using emission-line data, we can map the 37,363 sources on the WISE colour-colour plot in Figure \ref{fig_mpa} to define the loci of radio galaxies and SFGs. We note that SFGs are redder than radio galaxies in both $\rm W1-W2$ and $\rm W2-W3$ axes.

    In addition, SFGs generally exhibit lower radio luminosities compared to radio galaxies. The radio emission of the SFGs is primarily from regions associated with star formation, such as supernova remnants (SNRs) and HII regions \citep{2019sabeter}, and is less energetic than AGN jets and lobes. Most SFGs have $L_{144\mathrm{MHz}}<10^{25}~\mathrm{{W~Hz}^{-1}}$, and radio sources with luminosities higher than $ 10^{25}\ \mathrm{{W\ Hz}^{-1}} $ are more likely to be radio galaxies \citep[see e.g.,][]{gurkan2018,2019hardcastle}. We plot the luminous radio sources ($L_{144\mathrm{MHz}}>10^{25}~\mathrm{{W~Hz}^{-1}}$) as purple dots in Figure \ref{fig_mpa}, which are more likely to be radio galaxies than SFGs. However, 7\% of these luminous sources overlap with the region occupied by SFGs (W2 - W3 > 0.8). 
    We used the luminosity criteria provided by \cite{2019hardcastle} to avoid classifying luminous radio sources as SFGs. We classify sources as radio galaxies if they meet any of the following conditions: (1) they are outside the SFG region (W2 - W3 < 0.8); (2) they are within the SFG region and satisfy \( L_{144\mathrm{MHz}} > 10^{25}~\mathrm{W~Hz^{-1}} \) \& \(\mathrm{M_{Ks} > -25} \); or (3) they are within the SFG region and satisfy \(\mathrm{M_{Ks}} < -25 \) and \(\log_{10}(L_{144\mathrm{MHz}}) > 25.3 - 0.06(25 + \mathrm{M_{_s}}) \). Following this selection process, a total of 17,194 radio galaxies were identified.    
    
\begin{figure}[htb]
    \centering
    \includegraphics[width=0.38\textwidth]{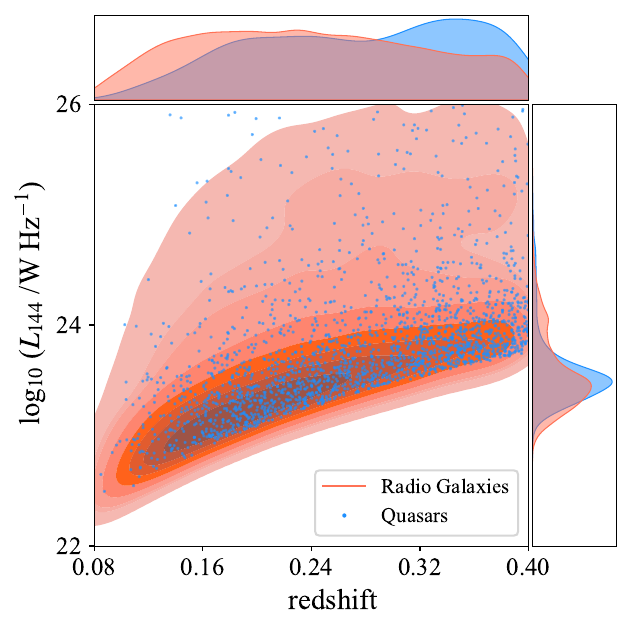}
    \caption{Rest-frame radio luminosity versus redshift of the selected radio galaxies and quasars. The density distribution of radio galaxies is shown as red contours, and quasars are shown as blue dots. The top and right panels feature the marginal density distributions of redshifts and radio luminosities for radio galaxies (red) and quasars (blue), respectively.}
    \label{fig_pz}
  \end{figure}

   Combining 9383 radio galaxies identified using emission-line data with 17,194 radio galaxies selected with WISE colours and radio luminosity, we obtained a total of 26,577 radio galaxies. Figure \ref{fig_pz} presents the rest-frame 144 MHz radio luminosity versus redshift for the 26,577 radio galaxies and 2028 quasars. Despite subtle differences, the radio galaxy and quasar samples are well-matched in both redshift and radio luminosity distributions. The median radio luminosity of radio galaxy sample is \(5.2 \times 10^{23}~\mathrm{{W~Hz}^{-1}} \), while the median radio luminosity of quasar sample is \(5.6 \times 10^{23}~\mathrm{{W~Hz}^{-1}} \).

   To enable a more detailed comparison between radio galaxies and quasars, we further classified the 9383 radio galaxies with emission-line data into HERGs and LERGs. Many studies have used emission line diagnostics to distinguish these two classes \citep[e.g.,][]{kewley2006,2010Buttiglione,2010cid,2012best&heckman}. In particular, \citet{2010Buttiglione} defined the Excitation Index (EI) using emission line flux ratios as: 
   
    \[
    EI = \log\left(\frac{\mathrm{[O~\textsc{iii}]}}{\mathrm{H{\beta}}}\right) - \frac{1}{3} \left[ \log\left(\frac{\mathrm{[N~\textsc{ii}]}}{\mathrm{H{\alpha}}}\right) + \log\left(\frac{\mathrm{[S~\textsc{ii}]}}{\mathrm{H{\alpha}}}\right) + \log\left(\frac{\mathrm{[O~\textsc{i}]}}{\mathrm{H{\alpha}}}\right)\right].
    \]
    \noindent Radio galaxies with \( EI \geq 0.95 \) are classified as HERGs, while those with \( EI < 0.95 \) are classified as LERGs. More recently, \citet{2024drake} developed a classification scheme that calculates the probability of different classes using the BPT diagram and Monte Carlo simulations, yielding highly reliable subclasses of radio galaxies while leaving some objects unclassified. To provide large samples of both HERGs and LERGs, we adopted the criteria on EI to classify the two subclasses.
    
\begin{figure*}[htb]
    \centering
       \includegraphics[width=0.75\textwidth]{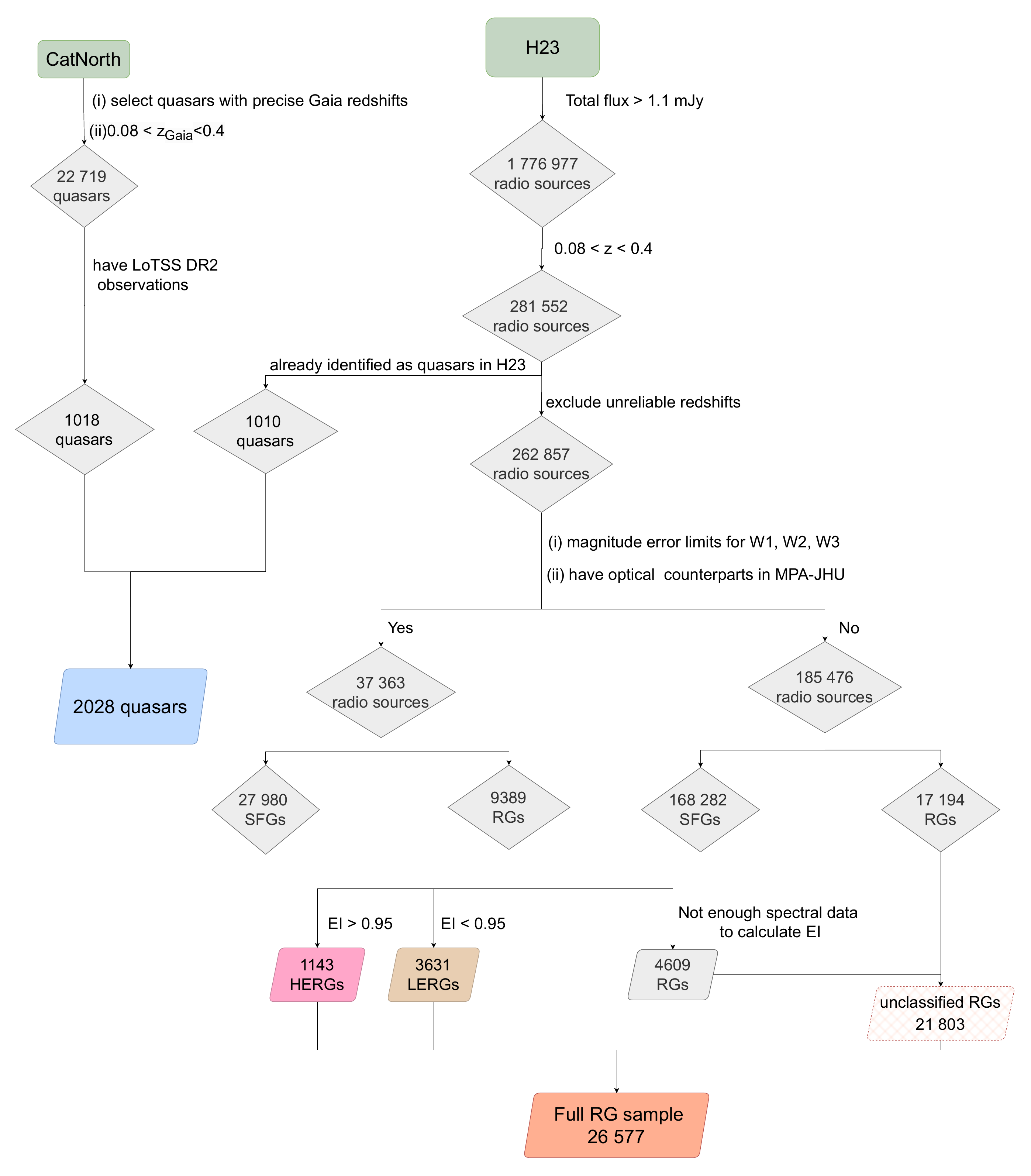}
       \caption{Flowchart showing the selection of samples of radio galaxies and quasars from H23. For those radio galaxies with spectral data but insufficient for classification and for those without spectral data, we refer to them as unclassified radio galaxies. These unclassified radio galaxies, along with HERGs and LERGs, make up our full radio galaxy sample.}
       \label{fig_enter-label}
   \end{figure*}

    Using EI calculated from the MPA-JHU catalog, we identify 1143 HERGs and 3631 LERGs. The HERG and LERG samples do not account for all 9383 radio galaxies with emission lines because only those sources with valid emission line measurements ($\mathrm{[O~\textsc{iii}]}$, $\mathrm{H{\beta}}$, $\mathrm{H{\alpha}}$, $\mathrm{[S~\textsc{ii}]}$, $\mathrm{[N~\textsc{ii}]}$, and $\mathrm{[O~\textsc{i}]}$) could be classified. In total, we provide three samples: the full sample of all 26,577 radio galaxies, along with the 1143 HERG sample and the 3631 LERG sample. Combined with the quasar sample, these samples enable a robust comparison between radio galaxies and quasars across different environments. Figure \ref{fig_enter-label} shows the schematic flowchart of the sample selection process.

\subsection{Crossmatching radio galaxies and quasars with the galaxy cluster catalogs}

     With the samples of radio galaxies and quasars established, we now proceed to examine their environmental properties by crossmatching them with the galaxy cluster catalogs. As suggested by \cite{2019croston}, an individual radio galaxy or quasar is considered to be associated with a galaxy cluster if the following criteria are satisfied: (i) the redshift difference $\Delta z$ between the galaxy and quasar and the cluster is less than 0.01; and (ii) the projected distance $\Delta D$ between the radio galaxy and quasar and the cluster is less than a matching radius of around 1 Mpc, for instance. We adopted these criteria, but used a larger matching radius of 2 Mpc, as quasars in clusters are often found in the outer regions at low redshifts. \citep{coldwell2006,2009lietzen}. We further justify the choice of the 2-Mpc radius below.

     The W12 catalog includes a measure of radius, $r_{200}$, for each cluster. Defined as the radius within which the average density is 200 times larger than the critical density of the Universe, $r_{200}$ is taken here as the boundary of a galaxy cluster. More than $40\%$ of clusters in W12 within $0.08 < z < 0.4$ have an $r_{200}$ larger than 1 Mpc, indicating that a radius of 1 Mpc does not cover the outskirts of many clusters. Due to the lack of radius information in R14, we extended the matching radius to a fixed 2 Mpc (99\% of clusters in W12 with $r_{200} < 2$ Mpc) to ensure a fair comparison between the two cluster catalogs.

     The thresholds of $\Delta D$ < 2 Mpc and $\Delta z$ < 0.01 are specifically designed for these two catalogs. The photometric uncertainties for these catalogs are approximately $\Delta z$ $\sim$ 0.014 (W12) and $\Delta z$ $\sim$ 0.006 (R14), making the  two criteria above suitable for our project. However, if future research focuses on exploring AGNs in lower-halo-mass clusters, the $\Delta D$ and $\Delta z$ thresholds may need to be adjusted to reflect the characteristics of such clusters.

\section{Results} \label{sec:result}

    Following the description of how we assembled the samples of radio galaxies, LERGs, HERGs, and quasars, then matched them with galaxy clusters within the defined search radius, in this section, we compare the trends seen in the match fractions for radio galaxies and quasars. We also consider their projected distances from the centers of associated clusters.

\subsection{The cluster match fractions of radio galaxies and quasars} \label{sec:result1}

    We present in Table \ref{tab:tab1} the numbers and fractions of radio galaxies and quasars associated with R14/W12 clusters within a 2-Mpc radius, namely, match numbers and match fractions. Given the significant difference in the number of the full radio galaxy sample compared to quasar+HERG+ LERG samples, we focus primarily on the association fractions. As shown in Table \ref{tab:tab1}, the cluster match fraction of all radio galaxies is more than twice that of quasars, for both R14 and W12 catalogs. 
  
    The cluster match fractions of LERGs and HERGs are also significantly higher than those for quasars. In particular, the match fraction of LERGs is more than four times higher than that of quasars. The higher match fractions with W12 in comparison to those with R14 also indicate higher completeness of W12. While we find that the results from W12 and R14 are consistent, we  mainly present those based on W12 hereafter for conciseness and clarity.

\begin{table*}[htb]
    \centering
    \caption{Match fraction with R14/W12 of quasars/radio galaxies within 2-Mpc}
    \label{tab:tab1}
    \resizebox{0.6\textwidth}{!}{%
    \begin{tabular}{ccccc}
    \hline
    \multicolumn{1}{l}{} & Quasars   & All RGs          & LERGs       & HERGs       \\ \hline
    R14                  & 37 ($1.8 \pm 0.3 \%$)& 1616 ($6.1 \pm 0.1 \%$)  & 175 ($4.8 \pm 0.3 \%$)  & 48 ($4.1 \pm 0.6 \%$)   \\
    W12                  & 84 ($4.1 \pm 0.4 \%$) & 4552 ($17.1 \pm 0.2 \%$) & 680 ($18.7 \pm 0.7 \%$) & 174 ($15.2 \pm 1.1 \%$) \\ \hline
    \end{tabular}
    }
\end{table*}

\begin{figure*}[h]
    \centering

    \begin{minipage}{0.32\textwidth} 
        \centering
        \includegraphics[width=\textwidth]{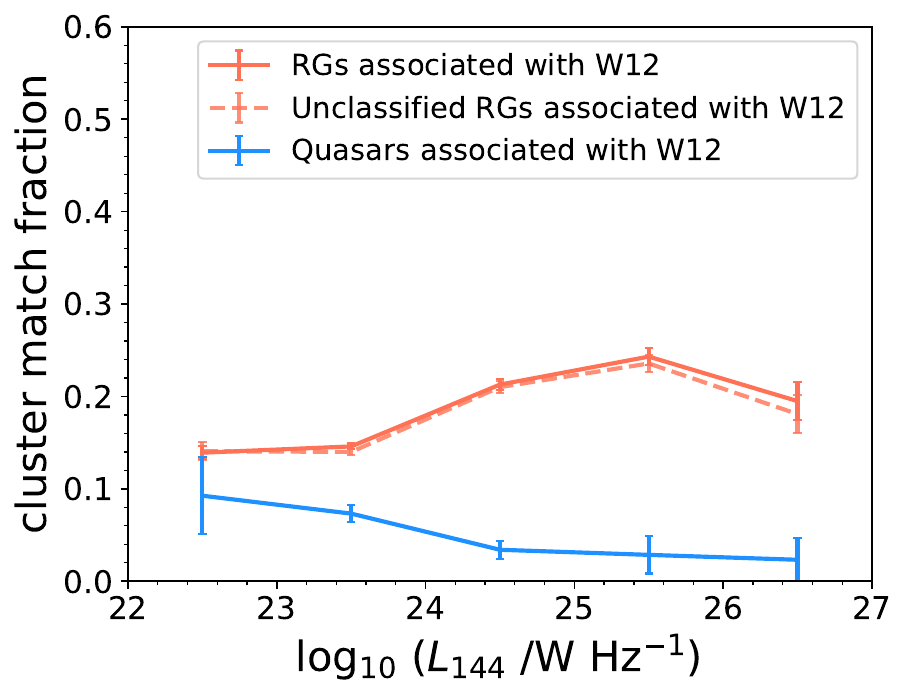}
        \subcaption{}
    \end{minipage}
    \hfill
    \begin{minipage}{0.32\textwidth}
        \centering
        \includegraphics[width=\textwidth]{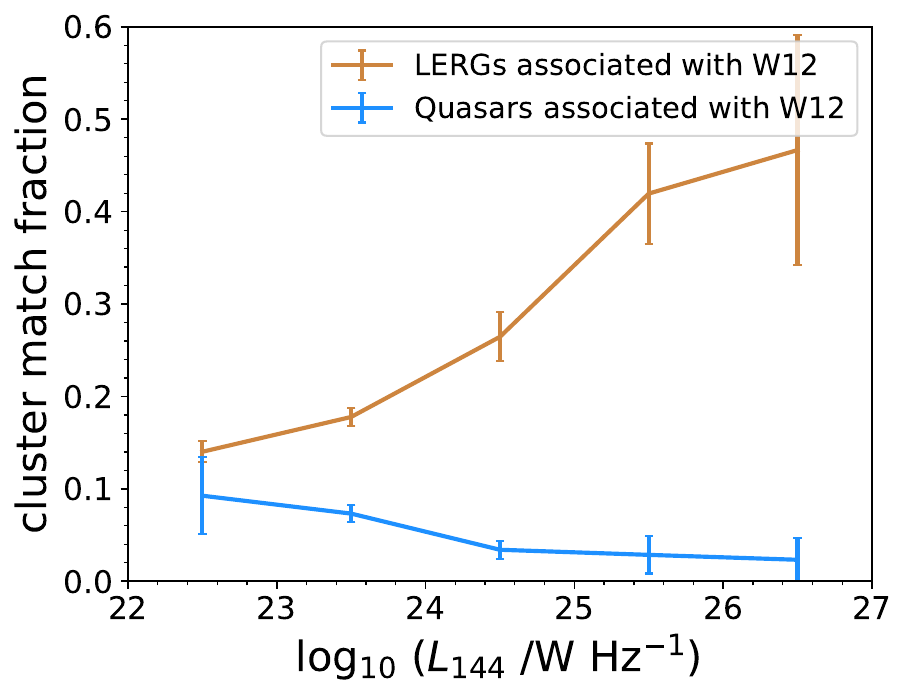}
        \subcaption{}
    \end{minipage}
    \hfill
    \begin{minipage}{0.32\textwidth}
        \centering
        \includegraphics[width=\textwidth]{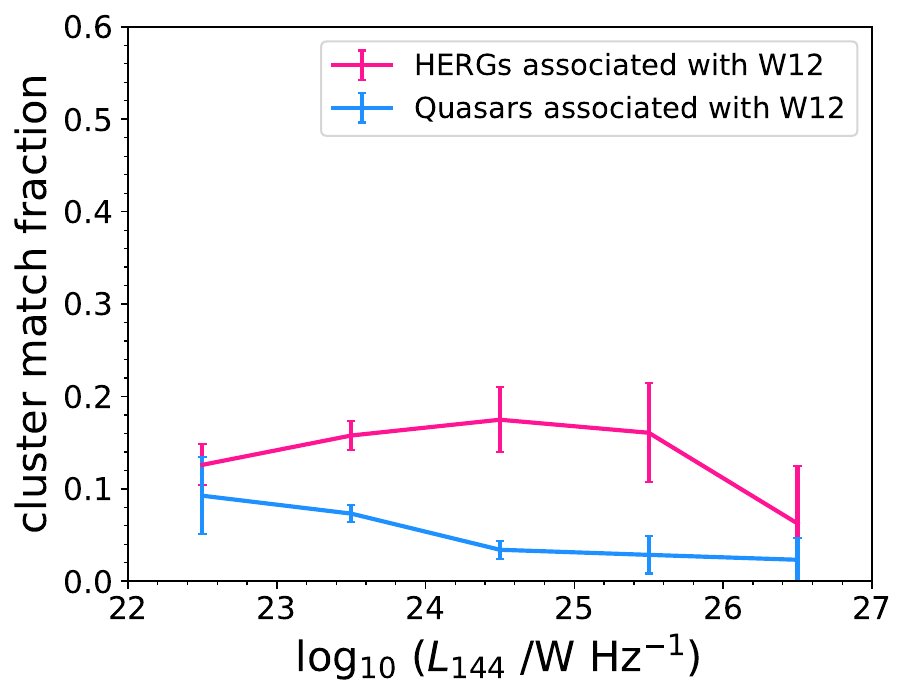}
        \subcaption{}
    \end{minipage}

    \caption{Relationship between the cluster match fraction and radio luminosity for radio galaxies and quasars within 2 Mpc. Panel (a) depicts the match fraction for all radio galaxies and quasars, with the orange dashed line representing the cluster match fraction for unclassified radio galaxies. Panel (b) shows the match fraction for LERGs and quasars and (c) is for HERGs and quasars. The blue lines in each panel represent the same quasar cluster match fraction but appear different due to the varying comparison groups. All error bars are calculated based on Poissonian distribution.}
    \label{fig_combined}

\end{figure*}

    We further investigate whether the match fractions depend on the 144 MHz radio luminosity. 
    Figure \ref{fig_combined} illustrates the relationship between the W12 cluster match fraction and the 144 MHz radio luminosity within a 2 Mpc radius. Panel (a) displays the match fractions for all radio galaxies, unclassified radio galaxies, and quasars within this radius. We include the cluster match fractions of unclassified radio galaxies, which show a trend similar to that of the full radio galaxy sample. This indicates that there is no fundamental difference between the classified samples (HERGs and LERGs) and the unclassified ones. Therefore, in the subsequent discussion, we focus only on the classified samples (HERGs and LERGs) and the full sample. As shown in panel (a), the cluster match fraction for all radio galaxies is, on average, three times that of quasars. The match fraction for all radio galaxies steadily increases with rising radio luminosity, although it experiences a slight decline at high luminosities ($L>10^{26}~\mathrm{W~Hz^{-1}}$). Overall, high-luminosity radio galaxies exhibit a higher cluster match fraction compared to low-luminosity ones. In contrast, for quasars, there is a downward trend in match fraction as radio luminosity increases, with high-luminosity quasars showing a match fraction of only \(2.3 \pm 2.3\)\%.

    Panel (b) illustrates the relationship between LERGs and quasars within search radii of 2 Mpc. On average, LERGs have a match fraction that is over four times that of quasars. Panel (b) shows that the match fraction for LERGs increases with radio luminosity, reaching up to \(47 \pm 12\)\% at higher luminosities. Panel (c) illustrates the relationship between HERGs and quasars within 2 Mpc. HERGs exhibit a higher match fraction than quasars. The match fraction trend for HERGs is similar to that of the full radio galaxy sample, showing a slight increase with rising radio luminosity below $10^{25}~\mathrm{W~Hz^{-1}}$. However, due to the small HERG sample size and large error bars, the trend at the high-luminosity end is less clear.

    Considering all panels in Figure \ref{fig_combined}, the match fractions of radio galaxies and their subclasses (LERGs and HERGs) consistently exceed those of quasars across all radio luminosity bins. Among these, LERGs exhibit the highest match fraction, while HERGs show a similar trend to the overall radio galaxy sample. Notably, the trends of all radio galaxies, LERGs, and HERGs all differ from those of quasars.

\subsection{The projected distance of radio galaxies/quasars to the center of galaxy clusters}

\begin{figure*}[htb]
    \centering
    \begin{minipage}{0.38\textwidth}
        \centering
        \includegraphics[width=\textwidth]{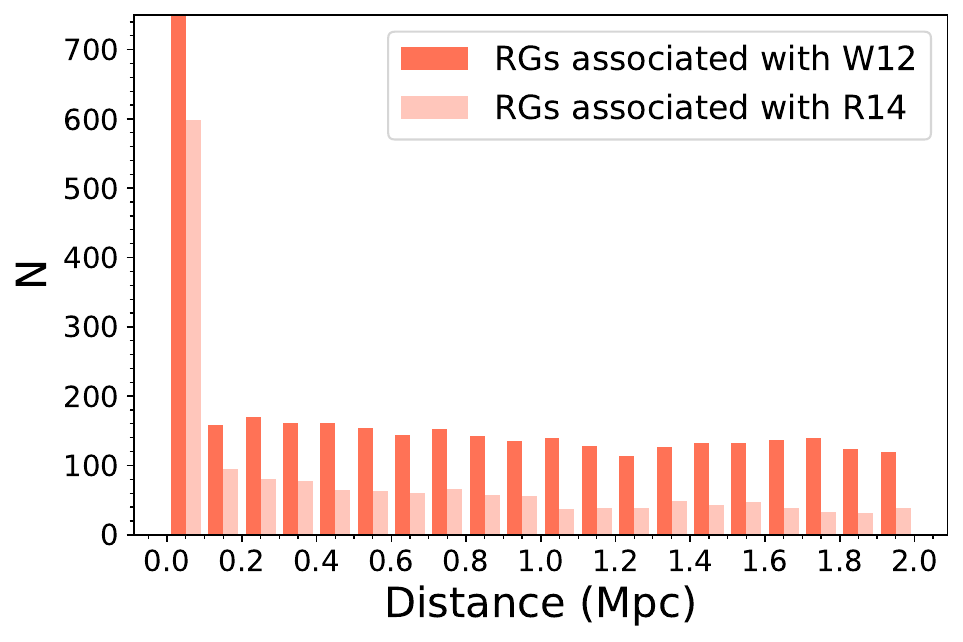}    
        \label{fig_prodis_rg}
        \subcaption[]{}
    \end{minipage}
    \begin{minipage}{0.37\textwidth}
        \centering
        \includegraphics[width=\textwidth]{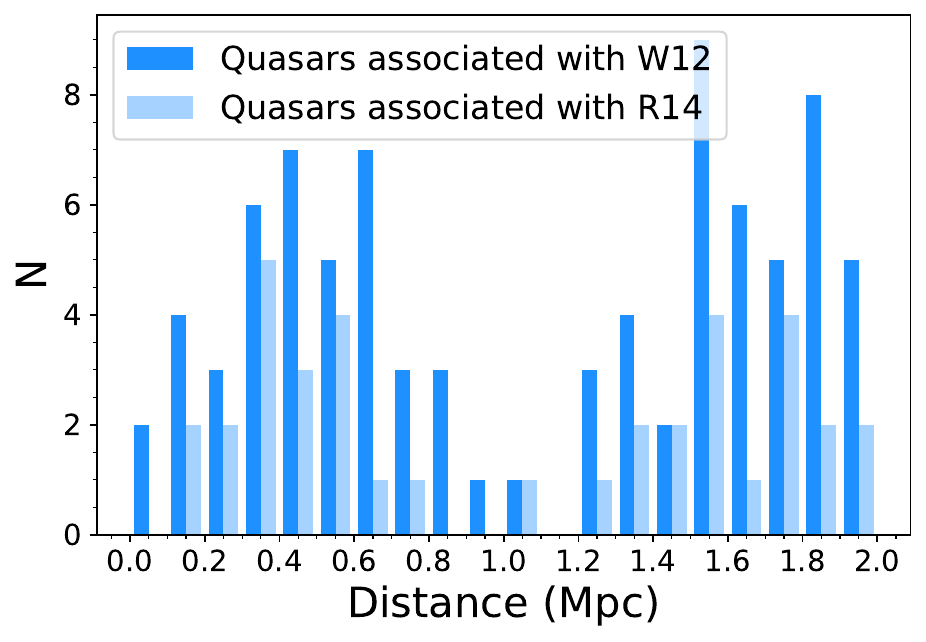}
        \label{fig_prodis_q}
        \subcaption[]{}
    \end{minipage}

    \begin{minipage}{0.38\textwidth}
        \centering
        \includegraphics[width=\textwidth]{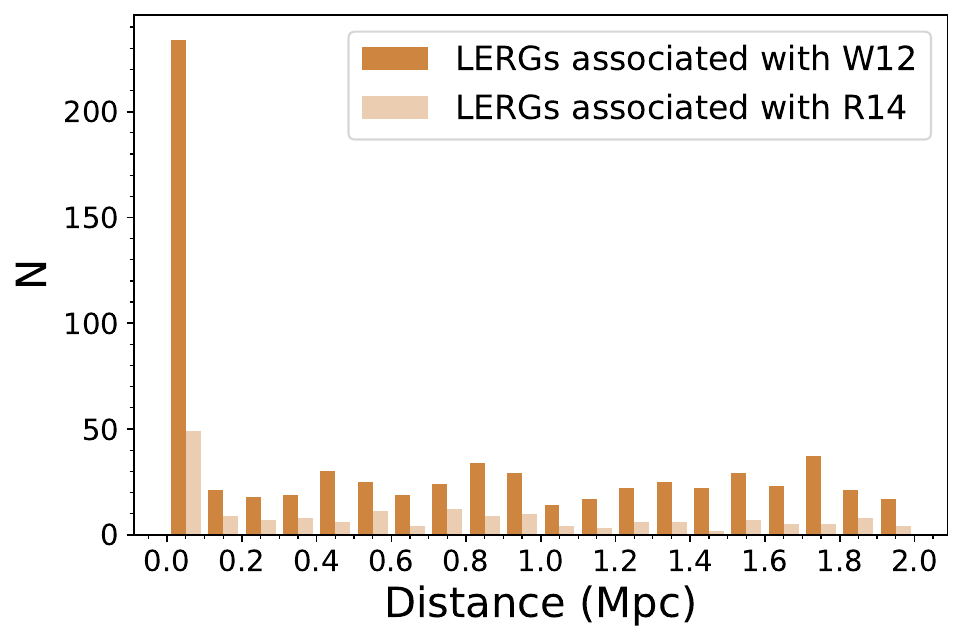}
        \label{fig_prodis_lerg}
        \subcaption[]{}
    \end{minipage}
    \label{fig_prodis}
    \begin{minipage}{0.38\textwidth}
        \centering
        \includegraphics[width=\textwidth]{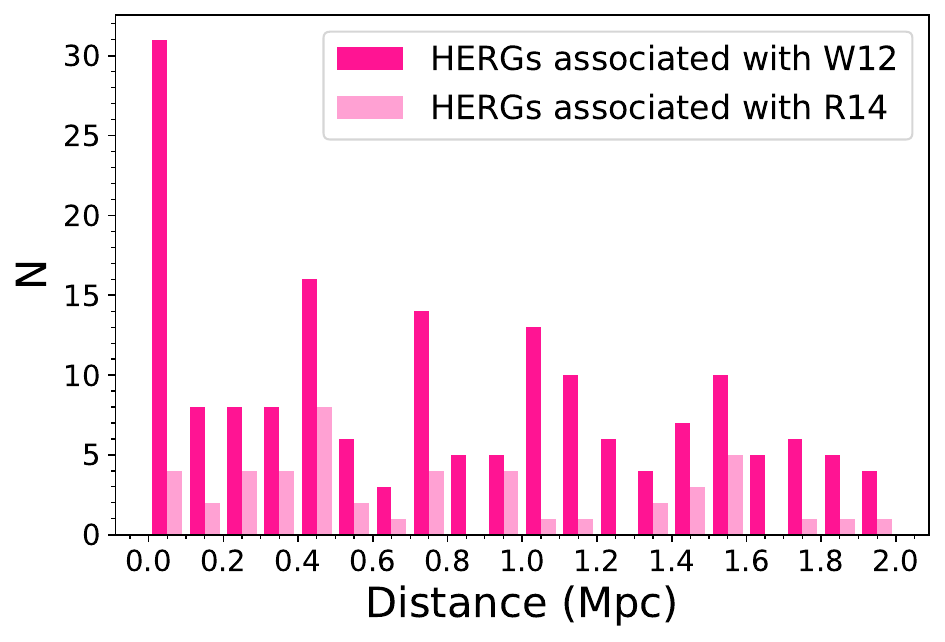}
        \label{fig_prodis_herg}
        \subcaption[]{}
    \end{minipage}
    \caption{Distribution of projected distances from radio galaxies/quasars to the centers of clusters. Most radio galaxies are in the very center of clusters. The y-axis of panel (a) has been cropped to make the distribution at large distances clearer. The number of radio galaxies located in the 0-0.1 Mpc range is 1882.}
    \label{fig_prodis}
\end{figure*}

    While the cluster match fractions within a fixed radius indicate significant differences between the environment of radio galaxies and that of quasars, their average number density profiles as a function of the projected distance to the centers of clusters can further probe the possible environmental preferences of each class of objects.
    
    In Figure \ref{fig_prodis}, we show the distribution of projected distances from the three radio galaxy samples and quasars to the centers of their associated clusters. Panel (a) clearly shows that, regardless of the choice of W12 or R14, most radio galaxies are within 100 kpc from the cluster centers. The median projected distance of radio galaxies and their associated W12 clusters is 337 kpc (336 kpc for R14). Panel (b) shows that quasars show no clear preference for locations within clusters. At a distance of 1 Mpc, the number of quasars drops to the lowest. The median projected distance of quasars and their associated W12 clusters is 1.12 Mpc (1.02 Mpc for R14). The trend shown in panel (c) is similar to that of radio galaxies in panel (a), with the largest number of LERGs associated with both R14 and W12 found within the 100 kpc range. The median projected distance of LERGs and their associated W12 clusters is 566 kpc (567 kpc for R14). Panel (d) shows that for HERGs associated with W12, the greatest number is also found within the central 100 kpc. However, for R14, due to the overall lower match numbers, there is no clear trend. The median projected distance of HERGs and their associated W12 clusters is 745 kpc (602 kpc for R14).

    Considering all panels in Figure \ref{fig_prodis}, radio galaxies and their subclasses (LERGs and HERGs) show median distances to their associated galaxy clusters that are consistently smaller than those of quasars. Although HERGs show less clear distribution trends due to fewer matched samples, the distance distribution trends of radio galaxies, LERGs, and HERGs all differ notably from those of quasars.

\section{Discussion}
    We examine the large-scale environments of 26,577 radio galaxies (1143 HERGs and 3631 LERGs) and 2028 quasars selected from the LoTSS DR2 optical catalog by crossmatching with SDSS group and cluster catalogs. The number of radio galaxies in our sample is about 13 times that of quasars. This could be due to the timescales over which quasars remain in their highly luminous state being shorter than the lifetimes of radio galaxies, which would have longer-lived radio emissions. In addition, some quasars are radio-quiet and would not be included in our quasar counts. These two reasons lead to a smaller number of radio-loud quasars compared to radio galaxies.
    
\subsection{The environmental properties of radio galaxies and quasars}

    We find that within a 2 Mpc matching radius, \(17.1\pm 0.2\% \) (\(6.1 \pm 0.1\)\%) of radio galaxies and \(4.1 \pm 0.3\)\% (\(1.8 \pm 0.3\)\%) of quasars are associated with clusters in the W12 (R14) catalogs. This is a more than twofold difference, which shows that radio galaxies have a significantly higher probability of residing in clusters than quasars. Both LERGs and HERGs show higher cluster match fractions compared to quasars. This reinforces the conclusion that radio galaxies are more likely to be found in clusters than quasars. Additionally, we find that LERGs have a higher match fraction than HERGs, suggesting that LERGs are more likely to be associated with clusters, consistent with the findings from \cite{2012best&heckman}. The match fractions for both quasars and radio galaxies in W12 are approximately three times those in R14, which is mainly attributable to the higher completeness of W12. Nevertheless, the overall trend in the match fraction for radio galaxies and quasars remains similar across both catalogs, irrespective of their completeness levels.

    When considering the variation of the match fraction with radio luminosity, we find that the cluster match fractions for all radio galaxies, LERGs, and HERGs in each luminosity bin are consistently higher than those for quasars; namely, at the same radio luminosity, quasars are less likely to be located in clusters. The match fractions for three radio galaxy samples increase with rising radio luminosity in the lower luminosity range, while the match fraction of quasars decreases with rising radio luminosity; namely,  luminous quasars are rare in clusters, while radio galaxies are not. We find that most radio galaxies and LERGs are close to the centers of galaxy clusters, whereas HERGs and quasars do not exhibit a clear positional preference.

\subsection{Main indications}
    Our main finding shows that radio galaxies are more likely to reside within clusters than quasars, contradicting the expectations of orientation-based unification.
    This difference is reflected in the properties of their host galaxies. Radio galaxies are typically hosted by massive elliptical galaxies with supermassive black holes \citep{2005best}, while the host galaxies of quasars can be spirals. Red elliptical galaxies are more commonly found in dense environments than blue spiral galaxies \citep{1997dressler,2010lee}. This pattern aligns with the evolution model by \cite{2006croton}, which suggests that radio galaxies may represent "radio mode" feedback, while quasars represent "quasar mode" feedback. According to this model, material accreting onto the central supermassive black hole in radio galaxies originates from gas inflow in the hot halo, which is abundantly available in such gas-rich environments as galaxy clusters \citep[see e.g.,][]{1988prestage,1991hill,2000worrall,2018Massaro,2019croston}. Studies based on 3CRR and NVSS samples are in line with this finding, showing that radio galaxies are usually located in denser environments, with LERGs tending to reside in denser environments than HERGs \citep[see e.g.,][]{2012best&heckman,2020massaro}. Our results confirm the findings of these studies. Conversely, quasars are more likely to be found in moderate environments with halo masses ranging from $10^{12}$ to $10^{13}~M_{\odot}$, which facilitate gas-rich mergers and interactions that are the primary triggering mechanisms for quasar activity \citep{2000kauffmann&haehnelt,2001Canalizo,2024breiding}.

\section{Conclusions}
    We present the largest available samples of LoTSS-DR2 radio galaxies and quasars at $0.08 < z < 0.4$ to test if the two classes of objects live in similar environments. The quasars are selected from the SDSS DR16Q and CatNorth catalogs, which provide reliable spectroscopic information from SDSS and Gaia. The radio galaxies are selected using optical emission line data and infrared WISE colours to minimize the contamination from SFGs. The radio galaxy and quasar samples are well-matched in redshift and luminosity, ensuring a robust comparison between the two samples. To further compare the environments of different types of radio galaxies with those of quasars, we categorized the radio galaxies with emission-line data into LERGs and HERGs. The difference between the environment of radio galaxies and their subclasses (LERGs and HERGs) and that of quasars challenges the orientation-based unification scheme. We find that:
    
    \begin{enumerate}
        \item The match fractions associated with clusters of radio galaxies ($17.1 \pm 0.2 \%$), LERGs ($18.7 \pm 0.7 \%$), and HERGs ($15.2 \pm 1.1 \%$) are higher than that of quasars ($4.1 \pm 0.4 \%$).
        \item Only $2.3 \pm 2.3\%$ luminous quasars are in clusters.
        \item In general, most radio galaxies are located near cluster centers while quasars do not show a strong preference for their positions in clusters.
    \end{enumerate}

    The environmental differences between radio galaxies and quasars are not in line with the unified model, but these differences may suggest the model’s limitations or indicate the need for further adjustments. 
    In this study, all W12 and R14 clusters with redshifts in the range  of $0.08 < z < 0.4$ have masses $M_{200} > 10^{14} \, M_{\odot}$, indicating that our results are related to massive clusters.
    Future studies could compare the similarities and differences between radio galaxies and quasars in moderate environments using number counts of galaxies \citep[see e.g.,][]{2008strand,2009lietzen}. In addition, the future WEAVE-LOFAR survey \citep{2016smith} will combine high-quality spectroscopic data from WEAVE with LOFAR’s sensitive radio observations to address key gaps in the understanding of radio galaxies and quasars. WEAVE’s precise redshift measurements will enable a detailed mapping of radio sources in three-dimensional space, essential for studying their distribution and relationships with cosmic structures. Additionally, the spectral data will reveal physical properties such as metallicity, ionization states, star formation rates, and AGN characteristics, clarifying the connection between radio emission, host galaxies, and environments. By linking LOFAR radio jet observations to intergalactic medium conditions, WEAVE-LOFAR will help explain how environments are linked to the evolution of these systems, offering a comprehensive framework for future studies.

\begin{acknowledgements}

TP acknowledges support from the CSC (China Scholarship Council)-Leiden University joint scholarship program. TP sincerely thanks Martin Hardcastle and Daniel Smith for their valuable suggestions for this paper. TP acknowledges the referee for critical and insightful remarks, which not only enhanced this work but also provided new perspectives for our future research. We make use of LOFAR, SDSS and Gaia data.

LOFAR data products were provided by the LOFAR Surveys Key Science project (LSKSP; \url{https://lofar-surveys.org/}) and were derived from observations with the International LOFAR Telescope (ILT). LOFAR (van Haarlem et al. 2013) is the Low Frequency Array designed and constructed by ASTRON. It has observing, data processing, and data storage facilities in several countries, which are owned by various parties (each with their own funding sources), and which are collectively operated by the ILT foundation under a joint scientific policy. The efforts of the LSKSP have benefited from funding from the European Research Council, NOVA, NWO, CNRS-INSU, the SURF Co-operative, the UK Science and Technology Funding Council and the Jülich Supercomputing Centre.

Funding for the Sloan Digital Sky Survey V has been provided by the Alfred P. Sloan Foundation, the Heising-Simons Foundation, the National Science Foundation, and the Participating Institutions. SDSS acknowledges support and resources from the Center for High-Performance Computing at the University of Utah. SDSS telescopes are located at Apache Point Observatory, funded by the Astrophysical Research Consortium and operated by New Mexico State University, and at Las Campanas Observatory, operated by the Carnegie Institution for Science. The SDSS web site is \url{www.sdss.org}.

SDSS is managed by the Astrophysical Research Consortium for the Participating Institutions of the SDSS Collaboration, including Caltech, The Carnegie Institution for Science, Chilean National Time Allocation Committee (CNTAC) ratified researchers, The Flatiron Institute, the Gotham Participation Group, Harvard University, Heidelberg University, The Johns Hopkins University, L’Ecole polytechnique f\'{e}d\'{e}rale de Lausanne (EPFL), Leibniz-Institut f\"{u}r Astrophysik Potsdam (AIP), Max-Planck-Institut f\"{u}r Astronomie (MPIA Heidelberg), Max-Planck-Institut f\"{u}r Extraterrestrische Physik (MPE), Nanjing University, National Astronomical Observatories of China (NAOC), New Mexico State University, The Ohio State University, Pennsylvania State University, Smithsonian Astrophysical Observatory, Space Telescope Science Institute (STScI), the Stellar Astrophysics Participation Group, Universidad Nacional Aut\'{o}noma de M\'{e}xico, University of Arizona, University of colourado Boulder, University of Illinois at Urbana-Champaign, University of Toronto, University of Utah, University of Virginia, Yale University, and Yunnan University.

This work has made use of data from the European Space Agency (ESA) mission
{\it Gaia} (\url{https://www.cosmos.esa.int/gaia}), processed by the {\it Gaia}
Data Processing and Analysis Consortium (DPAC,
\url{https://www.cosmos.esa.int/web/gaia/dpac/consortium}). Funding for the DPAC
has been provided by national institutions, in particular the institutions
participating in the {\it Gaia} Multilateral Agreement.
\end{acknowledgements}

%
%
\bibliographystyle{aa} 
\bibliography{sample}

\end{document}